\documentclass[preprint,aps,pre]{revtex4}
\usepackage{epsfig}
\usepackage{latexsym}

\begin{document}

\title{Directed Spiral Percolation Hull on the Square and Triangular
  Lattices}

\author{Santanu Sinha and S. B. Santra}

\affiliation{Department of Physics, Indian Institute of Technology
Guwahati, Guwahati-781039, Assam, India.}

\date \today

\vfill

\begin{abstract}
Critical properties of hulls of directed spiral percolation (DSP)
clusters are studied on the square and triangular lattices in two
dimensions ($2D$). The hull fractal dimension ($d_H$) and some of the
critical exponents associated with different moments of the hull size
distribution function of the anisotropic DSP clusters are reported
here. The values of $d_H$ and other critical exponents are found the
same within error bars on both the lattices. The universality of the
hull's critical exponents then holds true between the square and
triangular lattices in $2D$ unlike the cluster's critical exponents
which exhibit a breakdown of universality. The hull fractal dimension
$(d_H \approx 1.46)$ is also found close to $4/3$ and away from $7/4$,
that of ordinary percolation cluster hull. A new conjecture is
proposed for the hull fractal dimension ($d_H$) in terms of two
connectivity length exponents ($\nu_\|$ $\&$ $\nu_\perp$) of the
anisotropic clusters generated here. The values of $d_H$ and other
critical exponents obtained here are very close to that of the spiral
percolation cluster hull. The hull properties of the DSP clusters are
then mostly determined by the rotational constraint and almost
independent on the directional constraint present in the model.
\end{abstract}

\maketitle

\section{introduction}
Directed spiral percolation (DSP), a new site percolation model, is
recently introduced by Santra and Sinha \cite{dsp,sinha}. The DSP
model is constructed imposing both directional and rotational
constraints on the ordinary percolation model \cite{op}. The
directional constraint is in a fixed direction in space and the empty
sites in that direction are accessible to occupation. Due to the
rotational constraint the sites in the forward direction or in a
rotational direction, say clockwise, are accessible to occupation. The
direction of the rotational constraint is not fixed in space and it
depends on the direction from which the present site is occupied. The
cluster properties of the DSP model has already been studied on the
square\cite{dsp} and triangular\cite{sinha} lattices in two dimensions
($2D$) at their respective percolation thresholds. The DSP clusters
were found highly rarefied and anisotropic with chiral dangling
ends. It was observed that the values of the critical exponents
obtained on both the lattices are different from that of other
percolation models like ordinary percolation (OP)\cite{op}, directed
percolation (DP)\cite{dp}, and spiral percolation (SP)\cite{sp}. The
clusters on the triangular lattice were found more compact and less
anisotropic than the clusters on the square lattice due to higher
number of connectivity on the triangular lattice. It was found that
the DSP model not only belongs to a new universality class than other
percolation models but also exhibits a breakdown of universality in
the cluster properties between the square and triangular lattices in
$2D$\cite{sinha}.

The hull of a cluster is a continuous path of occupied sites at the
external boundary of the cluster. The hull has important significance
of its own apart from being a part of the percolation cluster. It has
connection with the physical problems like nucleation and
growth\cite{evan}, surface reaction\cite{sbsbs}, diffusion\cite{sapo}
etc. The hulls also exhibit scaling behaviour, different from that of
clusters, characterized by critical exponents at the percolation
threshold \cite{voss,bose}. Since the clusters generated in the DSP
model are anisotropic and chiral in nature, it is interesting to study
the critical behaviour of these anisotropic hulls with chiral dangling
ends at $p=p_c$.

In this paper, the critical properties of hulls of anisotropic DSP
clusters are studied on both the square and triangular lattices.
Results of anisotropic hull are not reported before and this is the
first study of anisotropic-chiral hull properties. A comparison is
made between the results obtained on the two lattices as well as in
different models, like OP and SP.

\section{Cluster Generation and Hull Extraction}

\begin{figure}
\centerline{\hfill  \psfig{file=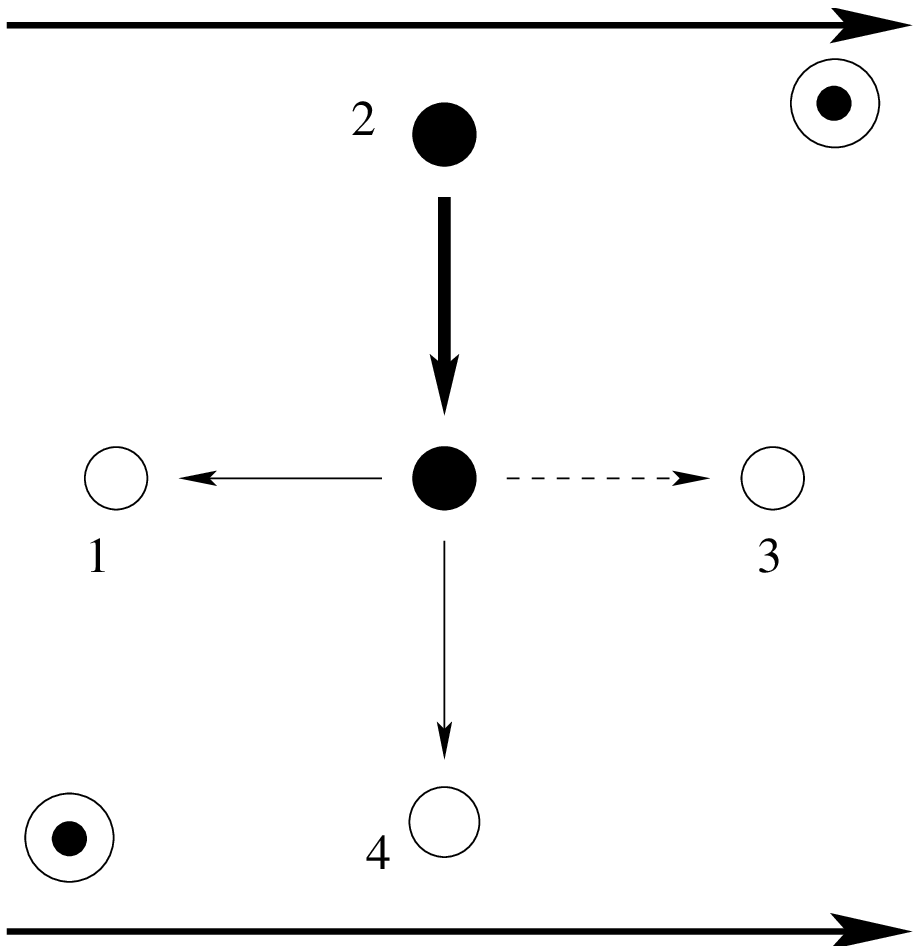,width=0.35\textwidth}
  \hfill \psfig{file=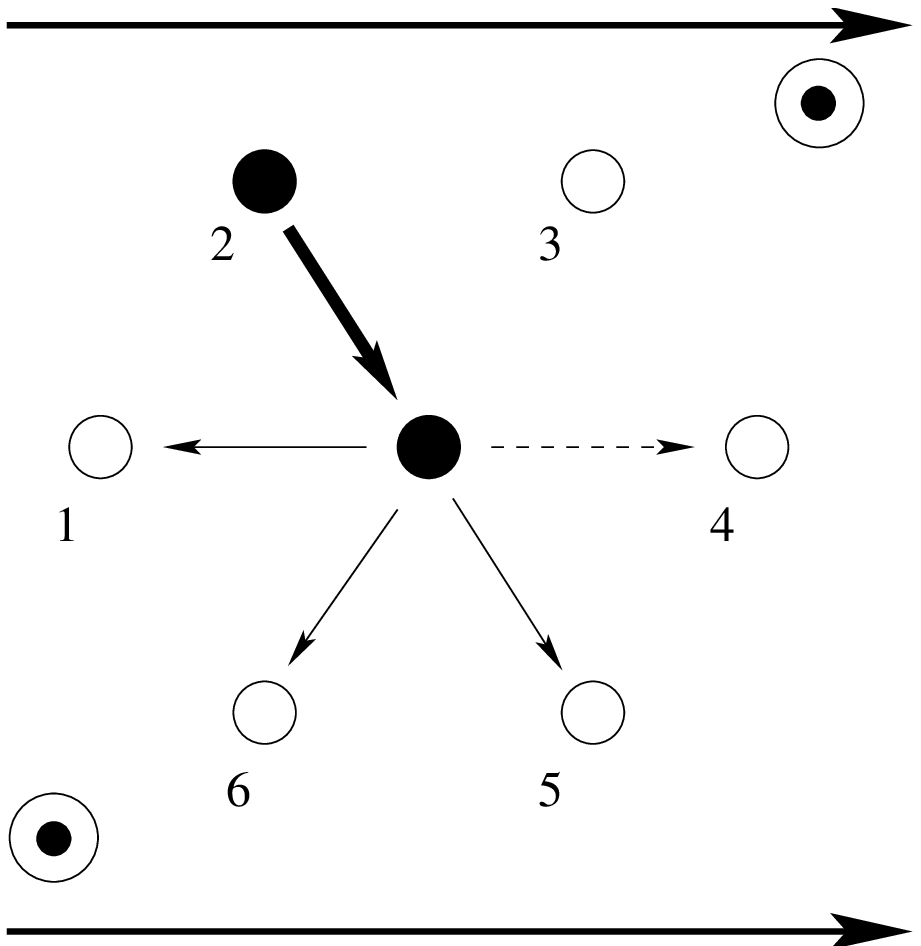,width=0.35\textwidth}\hfill}
\bigskip
\centerline{\hfill (a)\hfill\hfill  (b) \hfill}
\medskip
\caption{\label{demo}Selection of empty nearest neighbours for
occupation in a MC time step is demonstrated here on the square (a)
and triangular (b) lattices. Black circles are the occupied sites and
the open circles are the empty sites at a given time. Thick long
arrows from left to right represent the directional constraint. The
presence of clockwise rotational constraint is shown by encircled
dots. The central site is occupied from site 2 and shown by short
thick arrows. On the square lattice, site 3 on the right is always
eligible for occupation due to the directional constraint, marked by
dotted arrow and sites 4 and 1 are eligible for occupation due to the
rotational constraint, marked by thin arrows. Similarly on the
triangular lattice, site 4 due to the directional constraint and sites
5, 6 and 1 due to the rotational constraint are eligible for
occupation.}
\end{figure}

A single cluster growth Monte Carlo (MC) algorithm has already been
developed in Ref.\cite{dsp,sinha} to generate clusters under the
presence of both directional and rotational constraints following the
original algorithm of Leath\cite{leath}. A brief description of the
model is given here. In this algorithm, the central site of the
lattice is occupied with unit probability. All the nearest neighbours
of the central site can be occupied with equal probability $p$ in the
first time step. As soon as a site is occupied, the direction from
which it is occupied is assigned to it. Selection of empty nearest
neighbours in the next MC time steps is illustrated for the square
lattice in Figure \ref{demo}($a$) and triangular lattice in Figure
\ref{demo}($b$). The descriptions given below are valid for both
Figure \ref{demo}($a$) and \ref{demo}($b$). Two long arrows from left
to right represent the directional constraint. The presence of the
rotational constraint is shown by the encircled dots. The black
circles represent the occupied sites and the open circles represent
the empty sites. The direction from which the central site is occupied
is represented by a short thick arrow. The eligible empty site for
occupation due to the directional constraint is indicated by the
dotted arrow and the thin solid arrows indicate the eligible empty
sites for occupation due to the rotational constraint. There are two
empty sites due to the rotational constraint eligible for occupation
at any MC step on the square lattice, as shown in Figure
\ref{demo}($a$), whereas on the triangular lattice there are three
such sites available, as shown in Figure \ref{demo}($b$). After
selecting the eligible sites for occupation, they are occupied with
probability $p$. The coordinate of an occupied site in a cluster is
denoted by $(x$,$y)$. Periodic boundary conditions are applied in both
directions and the coordinates of the occupied sites are adjusted
accordingly whenever the boundary is crossed. At each time step the
span of the cluster in the $x$ and $y$ directions $L_x = x_{max} -
x_{min}$ and $L_y = y_{max} - y_{min}$ are determined. If $L_x$ or
$L_y\ge L$, the system size, then the cluster is considered to be a
spanning cluster. The critical percolation probability $p_c$ is
defined as below which there is no spanning cluster and at $p=p_c$ a
spanning cluster appears for the first time in the system.

As soon as a cluster is generated, the hull of the cluster is
determined by the method of Ziff {\em et al} \cite{ziff}. The hull of
a cluster is defined to be the continuous path of occupied sites at
the external boundary of the cluster. To determine the hull, a pair of
occupied and empty nearest neighbour sites on the external boundary is
chosen. An arrow is drawn from the empty site to the occupied site to
define a direction and one moves to the occupied site. Facing to the
direction of the arrow, a search is made starting from the left for an
occupied nearest neighbour. As soon as an occupied site is encountered
an arrow is drawn from the present site to the new occupied site and
one moves to the new occupied site. The process is continued until the
path passes the starting point in the same direction as it had first
been passed. All the occupied sites encountered during this walk are
listed in an array called as hull of the cluster. Typical hulls
extracted from the clusters generated at the percolation thresholds on
the square and triangular lattices are shown in Figure \ref{hullsq}
and Figure \ref{hulltr}, respectively. It could be seen from the hull
itself that the corresponding clusters are more anisotropic on the
square lattice than on the triangular lattice. However, both the hulls
contain chiral dangling ends and look very similar on small length
scales.

\section{Scaling Relations}
Analogous to the cluster related quantities, the hull related
quantities also exhibit critical properties at the percolation
threshold $p_c$. The corresponding critical exponents associated to
hull related quantities are expected to satisfy certain scaling
relations among themselves similar to the scaling relations satisfied
by the critical exponents of the cluster related quantities. The
scaling theory of the cluster related quantities for the DSP clusters
has already been developed in Ref.\cite{dsp}. In this section, the
critical exponents of the analogous hull related quantities will be
defined and their scaling relations will be developed. The size of a
hull $H$ is given by the number of occupied sites present on the hull
as the size of a cluster $S$ is given by the number of occupied sites
in a cluster. At $p_c$, the size of the infinite cluster goes as
\begin{equation}
\label{fd}
S_\infty \sim L^{d_f}
\end{equation}
with the system size $L$ and $d_f$ is the fractal dimension of the
infinite cluster. The value of $d_f$ has already been estimated on
both the square \cite{dsp} and triangular \cite{sinha} lattices. For
hulls, a similar relation is assumed here between the size of the
largest hull $H_\infty$ with the system size $L$ as
\begin{equation}
\label{hd}
H_\infty \sim L^{d_H}
\end{equation}
where $d_H$ is the hull dimension. The value of $d_H$ then can be
determined measuring $H_\infty$ for different system sizes $L$. It may
be noted here that the so-called Ziff's method \cite{ziff1} is not
suitable to determine the hull dimension of the anisotropic clusters
generated here. In Ziff's method, the linear distance is measured for
a given number of sites along the hull. Since there are two length
scales ($\xi_\|$ and $\xi_\perp$) involved in the DSP clusters, there
will be a crossover from $\xi_\perp$ to $\xi_\|$ as one measures the
linear distance from smaller number of sites to larger number of
sites. The average length of a given number of occupied sites will be
then strongly dependent on the initial point chosen.

From equations \ref{fd} and \ref{hd}, one can easily verify the
relation between the cluster size $S_\infty$ and the corresponding
hull size $H_\infty$ for the spanning clusters and it is given by
\begin{equation}
\label{fdhd}
H_\infty \sim S_\infty^x, \hspace{1cm} x=d_H/d_f
\end{equation}
where $d_H$ and $d_f$ are the fractal dimensions of the hull and the
corresponding percolation cluster, respectively. The value of $d_H$
obtained from equation \ref{hd} then could be verified measuring $x$ and
$d_f$ independently.

The hull size distribution is defined as
\begin{equation}
\label{hsd}
P_H(p)=N_H/N_{tot}
\end{equation}
where $N_H$ is the number of hulls of size $H$ and $N_{tot}$ is the
total number of hulls generated same as the number of clusters
generated. Analogous to the form of the cluster size distribution
function, the scaling function form of the hull size distribution is
assumed, as
\begin{equation}
\label{hsf}
P_H(p)=H^{-\tau_H+1}{\sf f}[H^{\sigma_H}(p-p_c)]
\end{equation}
where $\tau_H$ and $\sigma_H$ are two exponents. The same scaling
function form has already been verified for the spiral percolation
hull\cite{bose}.

Different moments of the hull size distribution $P_H(p)$, $\sum'_H
H^kP_H(p)$ are expected to be singular as $p\rightarrow p_c$. The
primed sum represents the sum of all finite hulls. The first, second
and third moments $\chi_H$, $\chi'_H$ and $\chi''_H$ of $P_H(p)$ are
calculated. The first moment $\chi_H=\sum'_HHP_H(p)$ is the average
hull size. The moments $\chi_H$, $\chi'_H$ and $\chi''_H$ diverge
with their respective critical exponents $\gamma_H$, $\delta_H$ and
$\eta_H$ at $p=p_c$. The critical exponents are defined as
\begin{equation}
\label{crte}
\chi_H \sim |p-p_c|^{-\gamma_H}, \hspace{0.2cm} \chi'_H \sim
|p-p_c|^{-\delta_H} \hspace{0.2cm} \& \hspace{0.2cm} \chi''_H \sim
|p-p_c|^{-\eta_H}.
\end{equation}

Since the hull related quantities are just different moments of the
hull size distribution function $P_H(p)$, the critical exponents
$\gamma_H$, $\delta_H$ and $\eta_H$ then should be related to $\tau_H$
and $\sigma_H$, exponents related to $P_H(p)$. It can be shown that
the $kth$ moment of the hull size distribution become singular as
\begin{equation}
\label{kmd}
\Sigma'_H H^k P_H(p) \sim (p-p_c)^{-(k-\tau_H+2)/\sigma_H} .
\end{equation}

The following scaling relations then can easily be obtained putting
appropriate values of $k$, order of the moment of $P_H(p)$, in equation
\ref{kmd} and one obtains,
\begin{equation}
\label{scr}
\gamma_H=(3-\tau_H)/\sigma_H, \hspace{0.2cm}
\delta_H=(4-\tau_H)/\sigma_H  \hspace{0.2cm} \&  \hspace{0.2cm}
\eta_H=(5-\tau_H)/\sigma_H .
\end{equation}

Eliminating $\tau_H$ and $\sigma_H$ from equation \ref{scr} a scaling
relation can be obtained as
\begin{equation}
\label{gde}
\eta_H=2\delta_H-\gamma_H .
\end{equation}
The exponents $\tau_H$ and $\sigma_H$ can also be estimated from the
measured exponents $\gamma_H$, $\delta_H$ and $\eta_H$ using the
following relations
\begin{equation}
\label{ts}
\begin{array}{l}
\tau_H=(3\delta_H -4\gamma_H)/(\delta_H-\gamma_H) =
(4\eta_H-5\delta_H)/(\eta_H-\delta_H) =
(3\eta_H-5\gamma_H)/(\eta_H-\gamma_H),\\ 

\sigma_H=1/(\delta_H-\gamma_H)=1/(\eta_H-\delta_H)=2/(\eta_H-\gamma_H).
\end{array}
\end{equation}

The values of the hull fractal dimension and other critical exponents
will be estimated below and the scaling theory will be verified.

\section{Results and Discussions}

The critical properties of the DSP clusters have already been studied
through Monte Carlo simulation and finite size scaling on both the
square and triangular lattices \cite{dsp,sinha}. The values of the
fractal dimension and other critical exponents of the cluster related
quantities were estimated numerically and are listed in table
\ref{table1}. The values of the critical exponents were found
different from that of the other percolation models like OP, DP and SP
and consequently the DSP model belongs to a new universality
class\cite{dsp}. It was observed that the spanning clusters are less
anisotropic and more compact on the triangular lattice than those on
the square lattice. This has happened due to the extra flexibility in
the spiraling constraint on the triangular lattice. The change in the
shape and compactness of the clusters had significant effect on the
values of some of the critical exponents and fractal dimension. They
were found considerably different on these two lattices for the DSP
model and leads to a breakdown of universality between the square and
triangular lattices in $2D$\cite{sinha}. Below, estimates of the hull
fractal dimension and other critical exponents of the hull related
quantities will be presented for anisotropic DSP cluster hulls for the
first time. A comparison will be made between the results obtained on
the square and triangular lattices, and also with that of the other
percolation models.

To extract hulls, clusters are generated both on the square and
triangular lattices at their percolation thresholds $p_{c} \approx
0.655$ and $0.570$ respectively. Different lattice sizes starting from
$L=2^5$ to $2^{11}$ have been used. Results are obtained through both
single lattice ($2^{11}\times 2^{11}$) Monte Carlo and finite size
scaling methods. A total number $N_{tot}$ of $5\times 10^4$ spanning
clusters are generated for each lattice size $L$ and every $p$, site
occupation prbability. Since each cluster has an associated hull, the
number of hulls extracted are also $N_{tot}$. The size of the hull $H$
is given by the number of occupied sites belonging to the hull.

\begin{figure}
\centerline{ \psfig{file=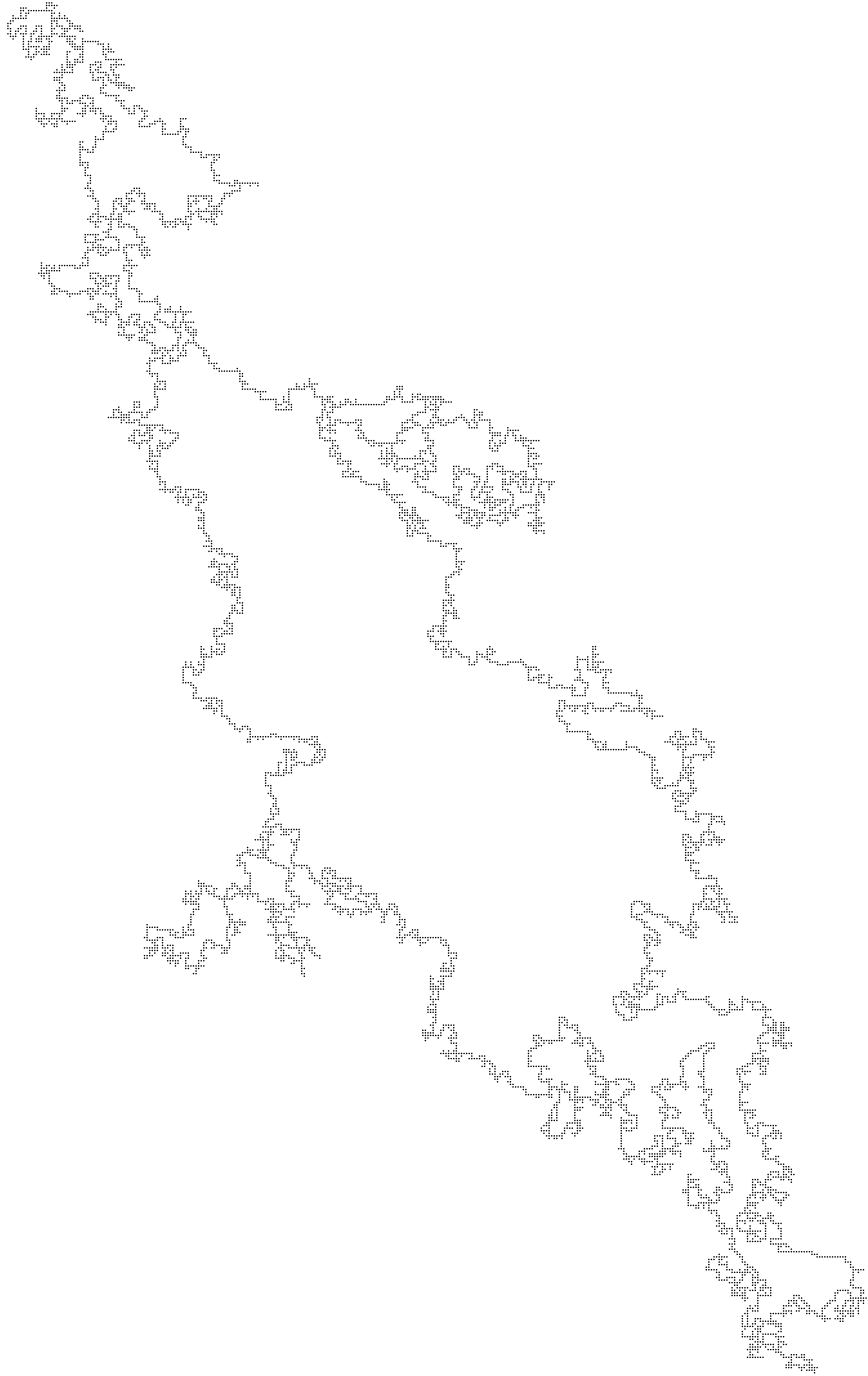,width=0.75\textwidth}} 
\bigskip
\caption{\label{hullsq} Hull of a spanning cluster of size $S=23964$
on a $256\times 256$ square lattice at the percolation threshold
$p_c=0.655$. The size of the hull is $H=7938$. It could be seen that
the dangling ends are clockwisely rotated and the cluster is
anisotropic.}
\end{figure}

\begin{figure}
\centerline{\psfig{file=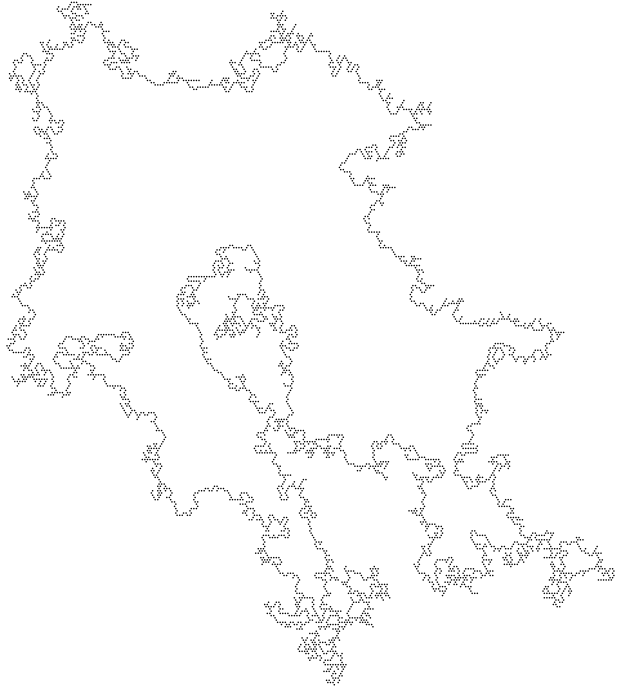,width=0.75\textwidth}}
\bigskip
\caption{\label{hulltr}Hull of a spanning cluster of size $S=15648$ on
a $256\times 256$ triangular lattice at the percolation threshold
$p_c=0.570$. The size of the hull is $H=4370$. There are chiral
dangling ends but the cluster is less anisotropic than the cluster on
the square lattice.}
\end{figure}

\begin{figure}
  \centerline{\psfig{file=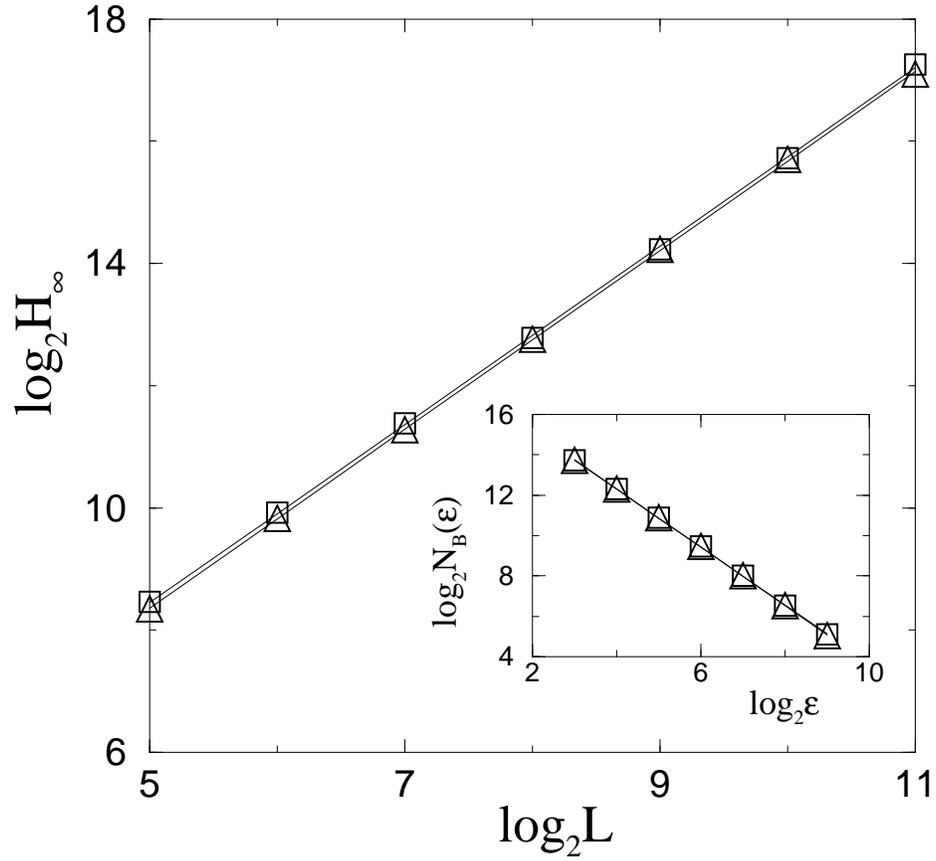,width=0.75\textwidth}}
  \bigskip
  \caption{\label{frdfsc} Average hull size $H_\infty$ of the spanning
  clusters versus the system size $L$. The squares represent the
  square lattice data and the triangles represent the triangular
  lattice data. The hull fractal dimensions are obtained as
  $d_H=1.458\pm 0.008$ for the square lattice and $1.463\pm 0.004$ for
  the triangular lattice. In the inset, the number of boxes
  $N_B(\epsilon)$ is plotted against the box size $\epsilon$. The hull
  fractal dimensions obtained by box counting method are $d_H=1.44\pm
  0.01$ for the square lattice and $1.45\pm 0.01$ for the triangular
  lattice. The values obtained in two different methods are within
  error bars. The value of $d_H$ is also within error bars on the two
  lattices. }
\end{figure}

The hull fractal dimension of the DSP clusters is determined following
the scaling relation, $H_\infty\sim L^{d_H}$, equation \ref{hd}. In
Figure \ref{frdfsc}, average hull size of the spanning clusters
$H_\infty$ is plotted against the system size $L$. The squares
represent the square lattice data and the triangles represent the
triangular lattice data. The values of the hull dimension $d_H$ are
obtained as $d_H = 1.458\pm 0.008$ for the square lattice and $d_H =
1.463\pm 0.004$ for the triangular lattice. The hull fractal dimension
$d_H$ has also been measured by box counting method generating
spanning clusters on the largest lattice ($2^{11}\times2^{11}$)
considered here. The number of boxes occupied with a hull site
$N_B(\epsilon) \sim \epsilon^{d_H}$, where $\epsilon$ is the box
size. The results are shown in the inset of Figure \ref{frdfsc}. The
values of $d_H$ obtained in the box counting method are given as $1.44
\pm 0.01$ for the square and $1.45 \pm 0.01$ for the triangular
lattice. The errors quoted for both the methods are the least square
fit errors taking into account the statistical error of each data
point. The results obtained through box counting method and finite
size scaling are within error bars. There are few things to
notice. First, the hull fractal dimensions $d_H$ measured on the
square and triangular lattices in $2D$ are found the same within error
bars, approximately $1.46$, whereas the cluster fractal dimension
$d_f$ were found different on the same lattices\cite{sinha}. It could
also be seen that, the hulls shown in Figure \ref{hullsq} and
\ref{hulltr} of the clusters generated on the square and triangular
lattices respectively are very identical on small length
scales. However, DSP clusters are more compact and less anisotropic on
the triangular lattice than on the square lattice. As a consequence,
the cluster fractal dimension $d_f$ and other critical exponents of
the DSP clusters were found different on these two lattices (see Table
\ref{table1}). Thus, the extra flexibility given on the rotational
constraint on the triangular lattice was only able to modify the
critical behaviour of the whole cluster but unable to modify the
critical properties of the external perimeter, the hull.

Second, the hull fractal dimension $d_H$ obtained here is smaller than
that of OP cluster hulls ($7/4$). The OP cluster hulls generally
contains long fjords or bays which are absent here. For OP cluster
hulls, the value of $d_H=7/4$ was conjectured by Sapoval {\em et al}
\cite{sapo} through a relation $d_H=1+1/\nu$ connecting the hull
dimension $d_H$ and the connectivity length exponent $\nu$ studying
the diffusion fronts. (In the case of OP, the value of $\nu$ is
$4/3$).  This prediction was supported through large scale simulation
by Ziff \cite{ziff1} and also proved analytically by Saleur and
Duplantier \cite{dup}. It has already been observed that the
conjecture does not hold true in the case of SP cluster hulls. The
values of the connectivity exponents and hull fractal dimensions
obtained for the SP clusters are: $\nu(SP) \approx 1.116$ and $d_H(SP)
\approx 1.476$ ($1+1/\nu \approx 1.896$) on the square lattice and
$\nu(SP) \approx 1.136$ and $d_H(SP) \approx 1.466$ ($1+1/\nu \approx
1.880$) on the triangular lattice \cite{bose}. Moreover, the DSP
clusters are anisotropic and there are two connectivity length
exponents $\nu_\parallel$ and $\nu_\perp$. The hull dimension $d_H$
here then may not be related to the connectivity length exponents
$\nu_\parallel$ and $\nu_\perp$ of the anisotropic clusters in a
simple manner as it was predicted by Sapoval {\em et al} for the case
of isotropic clusters generated in OP. Interestingly, it is found that
the hull dimension $d_H$ calculated form the following relation
\begin{equation}
\label{dhnu}
d_H=1+\frac{\nu_\perp}{\nu_\| + \nu_\perp}
\end{equation}
is very close to the numerical values obtained here. Using equation
\ref{dhnu}, the hull dimensions $d_H$ are obtained as $d_H\approx
1.46$ on the square lattice and $d_H\approx 1.47$ on the triangular
lattice. It seems that for the anisotropic clusters generated in DSP,
the hull dimension $d_H$ is connected to the connectivity length
exponents $\nu_\|$ and $\nu_\perp$ by the proposed relation in
equation \ref{dhnu} and the relation proposed by Sapoval {\em et al}
is valid only for the isotropic clusters generated in OP. However, an
argument similar to the one given by Sapoval {\em et al}\cite{sapo}
would be possible only from the study of ``anisotropic diffusion fronts''.

Third, the values of the hull fractal dimensions obtained for both the
DSP and SP clusters are close to the fractal dimension $D_e=4/3$ of
the externally accessible perimeter defined by Grossman and
Aharony\cite{gross}. The external perimeter defined by Grossman and
Aharony includes the sites available to a finite-size particle, coming
from the outside, that is touching the occupied sites on the
cluster. This external perimeter excludes deep fjords or bays and
consequently the fractal dimension found $4/3$. However, in the case
of SP and DSP clusters, the hull is compact (or smooth) in comparison
to the OP cluster hull due to the presence of rotational constraint in
these models. The rotational constraint produces compact chiral
dangling ends on the external boundary and makes the hull fractal
dimension $d_H$ closer to that of the Grossman-Aharony external
perimeter. It should be mentioned here that similar value of hull
fractal dimension close to $4/3$ was also obtained numerically by
Meakin and Family\cite{meakin} and Rosso\cite{rosso}. It was also
observed experimentally in the study of invasion percolation fronts by
Birovljev {\em et al}\cite{biro} and in the corrosion of thin Aluminum
film by Bal\'azs\cite{bala}. Deviation from the fractal dimension of
$7/4$ was also found in the study of self-stabilized etching of random
systems by Sapoval {\em et al}\cite{bssbs}. Recently, Aizenman {\em et
al} proved exactly the fractal dimension of Grossman-Aharony external
perimeter is $4/3$ \cite{dup1}.

\begin{figure}
  \centerline{ \psfig{file=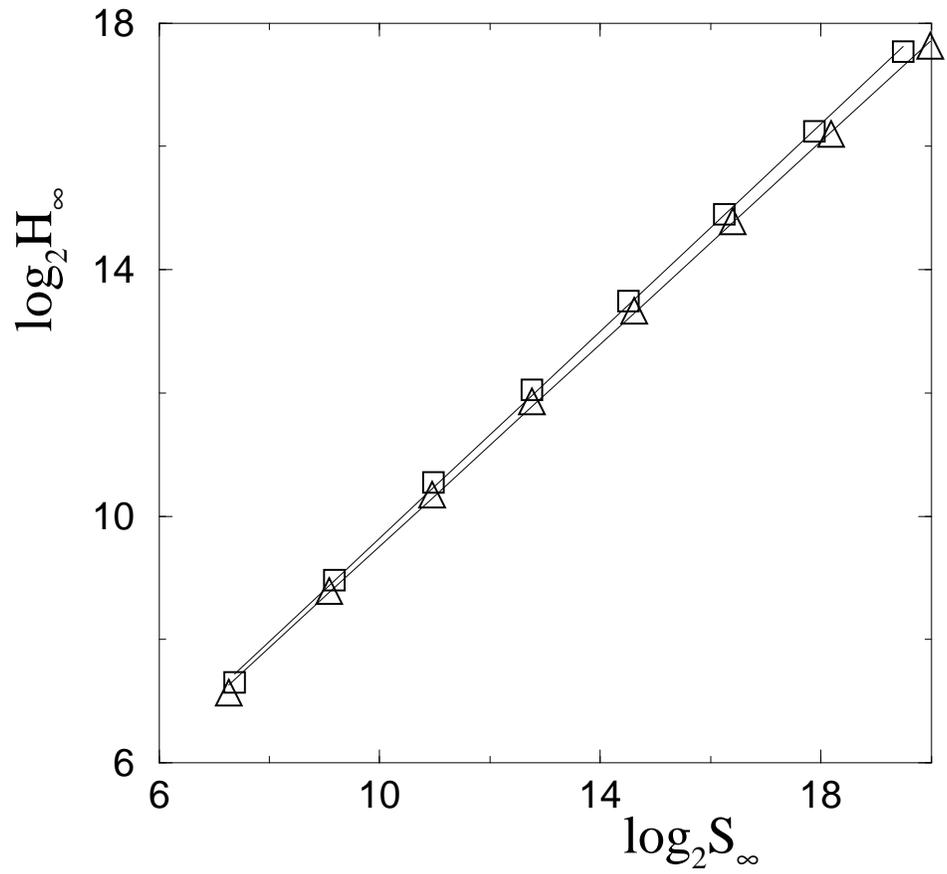,width=0.75\textwidth} }
  \bigskip
  \caption{\label{xsa} Plot of average hull size $H_\infty$ versus
  corresponding average cluster size $S_\infty$ of the spanning
  clusters. The squares represent the square lattice data and the
  triangles represent the triangular lattice data. The values of the
  exponent $x$ are obtained as $x=0.839\pm0.007$ for the square
  lattice and $x=0.820\pm0.006$ for the triangular lattice.}
\end{figure}

The relation between the hull dimension $d_H$ and the fractal
dimension $d_f$ of the percolation clusters is given by equation
\ref{fdhd}, $H_\infty \sim S_\infty^x, x=d_H/d_f$. This is verified
here and the exponent $x$ is calculated. In Figure \ref{xsa}, the
average hull size $H_\infty$ is plotted against the corresponding
average size of the spanning clusters $S_\infty$ for both the square
and triangular lattices. The slope $x=d_H/d_f$ is found as $x=0.839\pm
0.006$ on the square lattice and $x=0.820\pm 0.006$ on the triangular
lattice. The errors are least square fit error taking into account the
statistical error of each data point. The values of $x$ are found
close but only slightly different on the two lattices. This small
difference in the value of $x$ is consistent with the small difference
in the cluster fractal dimensions $d_f$ on the two lattices. The hull
dimension $d_H$ could be estimated from the relation $d_H=x\times d_f$
and compared with the measured values. Taking the values of $d_f$ from
Ref.\cite{dsp,sinha}, given in Table \ref{table1}, it is found that
$d_H = 1.45 \pm 0.01$ on the square lattice and $d_H = 1.46 \pm 0.01$
on the triangular lattice which are within error bars of the measured
value $\approx 1.46$. The errors quoted here are the propagation
errors.

\begin{figure}
  \centerline{ \psfig{file=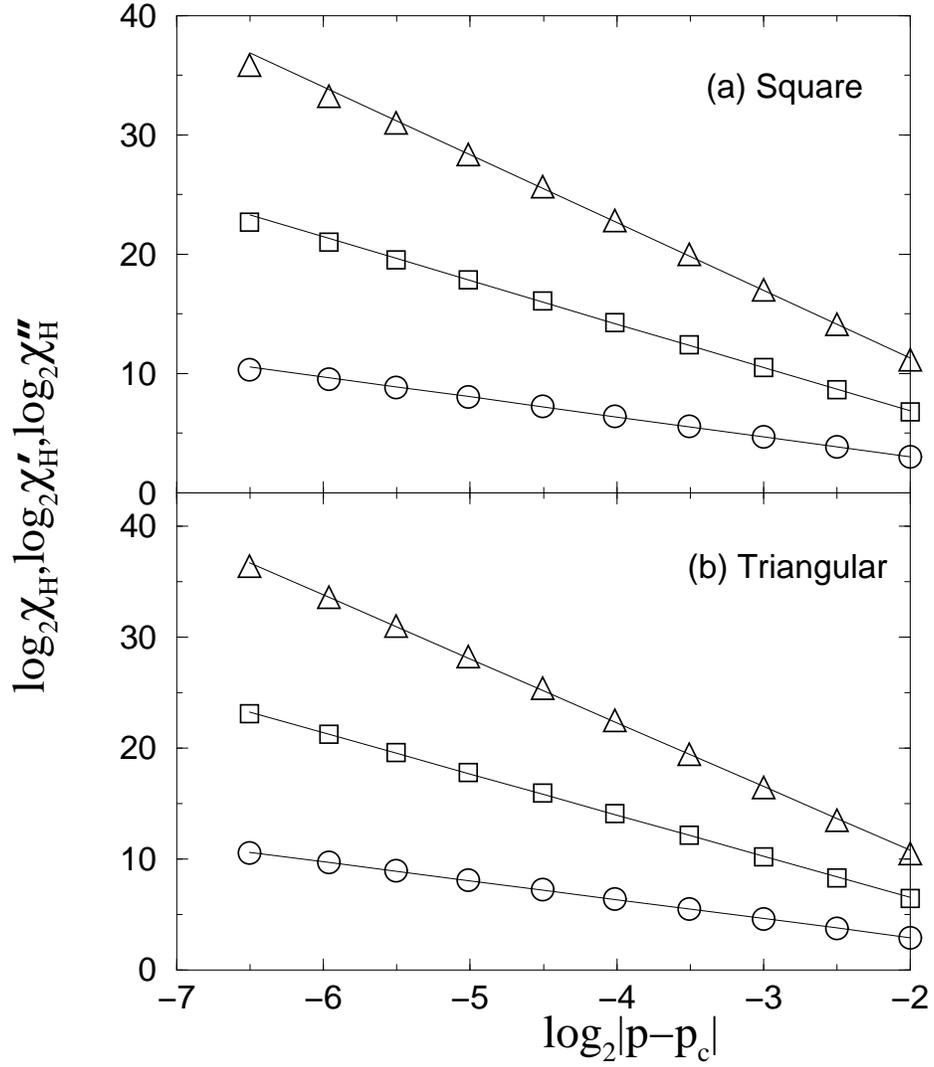,width=0.75\textwidth} }
  \bigskip
  \caption{\label{moments} Plot of the first, second and third moments
  $\chi_H$, $\chi'_H$ and $\chi''_H$ of hull size distribution versus
  $|p-p_c|$ for the lattice size $L=2048$ on the square $(a)$ and
  triangular $(b)$ lattices. Different symbols are circles for
  $\chi_H$, squares for $\chi'_H$ and triangles for $\chi''_H$ for
  each lattice. The solid lines represent the best fitted straight
  lines through the data points. The corresponding critical exponents
  are found as $\gamma_H=1.68\pm0.02$, $\delta_H=3.66\pm0.03$ and
  $\eta_H=5.70\pm0.05$ for the square lattice and $\gamma_H=1.71 \pm
  0.02$, $\delta_H=3.69\pm0.03$ and $\eta_H=5.73\pm0.05$ for the
  triangular lattice.}
\end{figure}

\begin{figure}
\centerline{\psfig{file=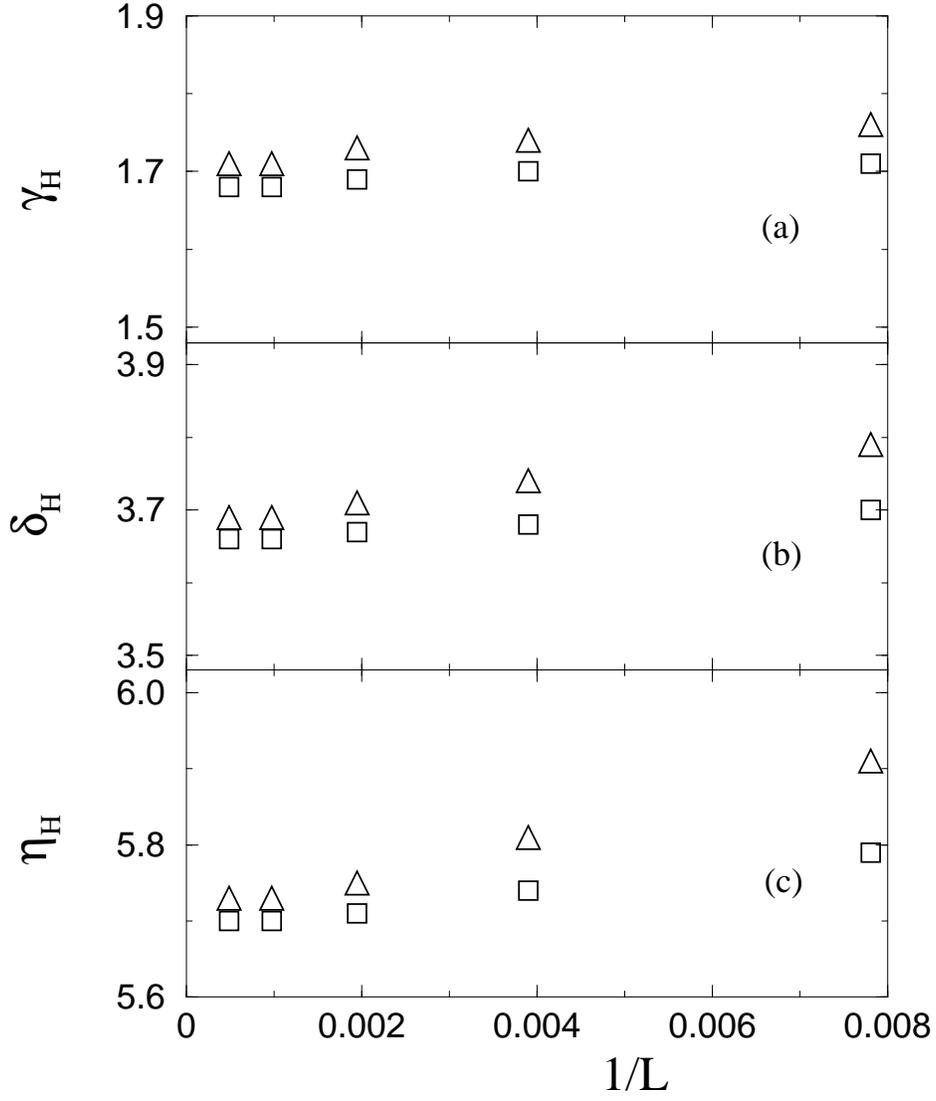,width=0.75\textwidth}}
\bigskip
\caption{\label{fscmoments} Plot of the hull moments exponents
$\gamma_H$ $(a)$, $\delta_H$ $(b)$and $\eta_H$ $(c)$ against the
inverse system size $1/L$. The system sizes considered are: $L= 128,
256, 512, 1024$ and $2048$. The squares represent square lattice data
and the triangles represent triangular lattice data. As $L\rightarrow
\infty$, the exponents are converging to the values corresponding to
$L=2048$. The values of the exponents are within the error bar on the
two lattices.}
\end{figure}

Now, the critical properties of different moments of the hull size
distribution function are studied. The critical exponents related to
the different moments of the hull size distribution $P_H(p)$ are
already defined in section III and the scaling relations are also
described there. In this section, the values of the critical exponents
are estimated and the scaling relations are verified.  The first three
moments $\chi_H$, $\chi'_H$ and $\chi''_H$ are plotted against
$|p-p_c|$ for the square lattice in Figure \ref{moments}($a$) and for
the triangular lattice in Figure \ref{moments}($b$). The system size
is taken as $L=2048$. The circles represent the average hull size
$\chi_H$, the squares represent $\chi'_H$, the second moment and the
triangles represent $\chi''_H$, the third moment in both the plots.
The values of the exponents obtained are $\gamma_H=1.68 \pm 0.02$,
$\delta_H=3.66\pm 0.03$ and $\eta_H=5.70\pm 0.05$ for the square
lattice. For the triangular lattice, the values of the exponents
obtained are $\gamma_H=1.71 \pm 0.02$, $\delta_H = 3.69\pm 0.03$ and
$\eta_H=5.73\pm 0.05$. The errors quoted here are the standard least
square fit error taking into account the statistical error of each
single data point. The system size $L$ dependence of the values of the
critical exponents of different moments of the hull size distribution
on the square and triangular lattices is also studied measuring the
exponents' values at different $L$. The values of $\gamma_H$,
$\delta_H$ and $\eta_H$ are plotted against the inverse system size
$1/L$ in Figure \ref{fscmoments}$(a)$, $(b)$ and $(c)$
respectively. The squares represent the square lattice data and the
triangles represent the triangular lattice data in each plot. The
values of the exponents obtained on these two lattices are converging
to the values of the exponents obtained for $L=2048$ through MC
simulation. The values of the hull moment exponents are also within
the error bars on the square and triangular lattices. The hull fractal
dimension $d_H$ is already found the same on these two lattices. It
should be mentioned here that the values of the critical exponents of
the analogous cluster related quantities were found different on the
square and triangular lattices \cite{sinha}. Thus, breakdown of
universality occurs in the cluster properties whereas universality
holds for the associated hull properties in the DSP model in $2D$
between the square and triangular lattices. This is a new observation
and appears for the first time in a percolation model. The scaling
relations could be verified now. The value of $2\delta_H - \gamma_H =
5.64\pm 0.08$ is very close to the value of the exponent $\eta_H
\approx 5.70$ for the square lattice. For the triangular lattice,
$2\delta_H - \gamma_H = 5.67 \pm 0.08$ and $\eta_H \approx 5.73$. The
scaling relation $\eta_H = 2\delta_H - \gamma_H$ (equation \ref{gde})
is then valid within the error bars on both the lattices.

In table \ref{table2}, the values of the hull fractal dimension and
other critical exponents are summarized and a comparison of the
results obtained on the square and triangular lattices as well as on
the SP model has been made. It can be seen that the universality holds
for the hull results on the square and triangular lattice, as it is
already mentioned, though there is a breakdown of universality for the
cluster properties on the same lattices. It is interesting to note
that the values of the exponents obtained here, especially the hull
fractal dimension $d_H$, are very close to that of the SP (percolation
under rotational constraint only) clusters hull\cite{bose} on both the
lattices. However, the SP clusters were found much more compact than
the DSP clusters and are isotropic. The fractal dimension $d_f$ of SP
clusters on the square and triangular lattices are $\approx 1.957$ and
$1.969$ respectively\cite{bose} whereas that of the DSP clusters are
$\approx 1.733$ \cite{dsp} and $1.775$ \cite{sinha} respectively. It
can be seen from the hull structures, given in Figure \ref{hullsq} and
\ref{hulltr}, that there exist chiral (clockwisely rotated) dangling
ends on the perimeter as in the case of SP cluster hulls. The hull
properties of both the DSP and SP clusters are then determined by the
existence of chiral dangling ends on the external perimeter. These
chiral dangling ends are generated due to the presence of the
rotational constraint in both the models. The directional constraint
then has a little effect on the hull critical properties and it is
mostly determined by the rotational constraint present in the
model. In a physical situation like transport of classical charged
particles in disordered systems in the presence of crossed electric
and magnetic fields, the magnetic properties then could be extracted
from the external perimeter only whereas the electrical properties
could be obtained from the bulk of the cluster.

\begin{figure}
\centerline{\psfig{file=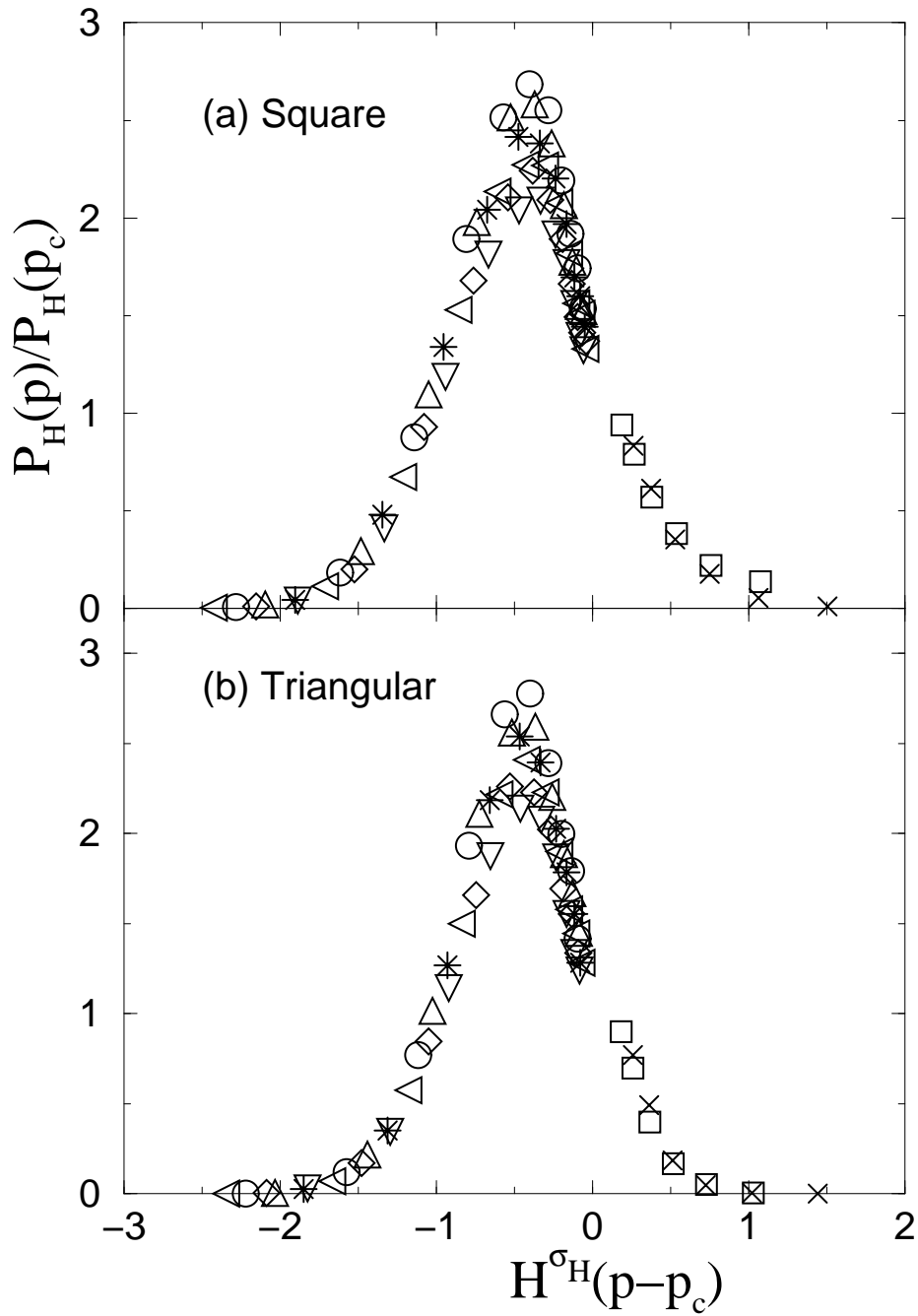,width=0.75\textwidth}}
\bigskip
\caption{\label{datacol} Plot of scaled hull size distribution
$P_H(p)/P_H(p_c)$ versus scaled variable $H^{\sigma_H}(p-p_c)$ for
different values of $p$ on the square $(a)$ and triangular $(b)$
lattices. The value of $\sigma_H$ is taken as $\sigma_H=0.498$
(square) and $\sigma_H=0.497$ (triangular). The hull size $H$ changes
from $64$ to $16384$. The data plotted correspond to $p-p_c= 0.007
(\times)$, $0.005 (\Box)$, $-0.035 (\nabla)$, $-0.04 (\Diamond)$,
$-0.045 (\triangleleft)$, $-0.05 (\ast)$, $-0.055 (\bigtriangleup)$,
$-0.06 (\bigcirc)$ for both the plots. Reasonable data collapse are
observed on both the lattices. Since the value $\sigma_H$ is almost
the same, the scaling function form looks almost identical on the two
lattices.}
\end{figure}

Finally, the form of the hull size scaling function $P_{H}(p) =
H^{-\tau_H+1} {\sf f}[H^{\sigma_H}(p-p_{c})]$ is verified. The values
of the exponents $\tau_H$ and $\sigma_H$ have been estimated using the
scaling relation given in equation \ref{ts}. For the square lattice,
the estimates of $\tau_H$ and $\sigma_H$ are obtained as
$\tau_H=2.17\pm 0.02$ and $\sigma_H=0.498\pm 0.003$. These values are
within the error bars of the values of $\tau_H$ and $\sigma_H$
obtained on the triangular lattice as $\tau_H=2.16 \pm 0.02$ and
$\sigma_H=0.497 \pm 0.003$. The errors mentioned correspond to
propagation error only. The scaling function form is verified through
data collapse by plotting $P_H(p)/P_H(p_c)$ against the scaled
variable $H^{\sigma_H}(p-p_c)$ for both the square and triangular
lattices in Figure \ref{datacol}$(a)$ and $(b)$, respectively. The hull
size varies from $64$ to $16384$ and $(p-p_c)$ is in the range $0.007$
to $-0.06$. Not only a reasonable data collapse is observed for both
the lattices but also the scaling function forms are found identical
on the two lattices.

\section{Conclusion}

Hulls of the DSP clusters are extracted generating large clusters at
the percolation threshold and their properties are analyzed on both
the square and triangular lattices in $2D$. Results are compared with
that of other percolation models. The values of the hull fractal
dimension $d_H$ and critical exponents $\gamma_H$, $\delta_H$ and
$\eta_H$ related to the hull size distribution function $P_H(p)$ are
determined. It is found that the values obtained for the hull fractal
dimension and the critical exponents related to the hull size
distribution are the same within error bars on the square and
triangular lattices unlike the cluster fractal dimension $d_f$ and
cluster related critical exponents of the DSP model. Thus, the
universality of the hull critical exponents holds for the square and
triangular lattices in $2D$ whereas there is a breakdown of
universality in the corresponding cluster properties. The value of
$d_H \approx 1.46$ obtained here is close to $4/3$, the fractal
dimension of the Grossman-Aharony external perimeter and away from
$7/4$ proposed by Sapoval {\em et al} for the ordinary percolation
hull. A new conjecture is made relating $d_H$ and two connectivity
length exponents $\nu_\|$ and $\nu_\perp$ for the anisotropic
clusters. It is found that the values of $d_H$ and other hull critical
exponents are close to that of the SP cluster hulls. Hull critical
properties are then mostly determined by the rotational constraint
than the directional constraint present in the model. The critical
exponents estimated here satisfy the scaling relations among
themselves within error bars. The assumed scaling function form has
been verified and found similar on both the square and triangular
lattices.

\section{acknowledgment} The authors thank A. Srinivasan for
helpful discussions. SS thanks CSIR, India for financial support.

\newpage

\begin{table}
\begin{tabular}{p{3.0cm}p{1.7cm}p{1.4cm}p{1.4cm}p{1.4cm}p{1.4cm}p{1.4cm}p{1.4cm}p{1.2cm}}
    \hline

    Lattice Type & $d_f$ & $\gamma$ & $\delta$ & $\eta$ &
    $\nu_\parallel$ & $\nu_\perp$ & $\sigma$ & $\tau$\\
    
    \hline Square\cite{dsp}: & $1.733$ & $1.85$ & $4.01$ & $6.21$ &
    $1.33$ & $1.12$ & $0.459$ & $2.16$\\& $\pm0.005$ & $\pm 0.01$ &
    $\pm 0.04$ & $\pm 0.08$ & $\pm 0.01$ & $\pm 0.03$ & $\pm 0.015$ &
    $\pm 0.20$ \\
    \begin{tabular}{p{3.0cm}p{2.6cm}}
     & $1.72\pm0.02$(FS)
    \end{tabular}
    \\ Triangular\cite{sinha}: & $1.775$ & $1.98$ & $4.30$ & $6.66$ &
    $1.36$ & $1.23$ & $0.427$ & $2.16$\\ & $\pm0.004$ & $\pm 0.02$ &
    $\pm 0.04$ & $\pm 0.08$ & $\pm 0.02$ & $\pm 0.02$ & $\pm 0.003$ &
    $\pm 0.02$\\
   \begin{tabular}{p{3.0cm}p{2.6cm}}
     & $1.80\pm0.03$(FS)
   \end{tabular}
    \\
    \hline
\end{tabular}
\bigskip
\caption{\label{table1} Numerical estimates of the critical exponents
 and fractal dimension measured for the DSP clusters on the square and
 triangular lattices. Some of the critical exponents and the fractal
 dimension are significantly different on the two lattices.}
\end{table}

\vfill
\begin{table}
\begin{tabular}{p{3.0cm}p{1.5cm}p{1.4cm}p{1.4cm}p{1.4cm}p{2.0cm}p{1.4cm}p{1.4cm}p{1.4cm}}   
    \hline Lattice Type & Model & $d_H$ & $d_H^c$ & $x$ & $d_H=xd_f$ &
    $\gamma_H$ & $\delta_H$ & $\eta_H$ \\
    \hline
 
    Square & DSP & $1.458$ & $1.46$ & $0.839$ & $1.45$ & $1.68$ &
    $3.66$ & $5.70$ \\
 
    & & $\pm 0.008$ & $\pm 0.02$ & $\pm 0.007$ & $\pm0.01$ & $\pm 0.02$
    & $\pm 0.03$ & $\pm 0.05$ \\
    \begin{tabular}{p{3.0cm}p{1.5cm}p{8.0cm}}
       && $d_H=1.44\pm0.01$ (Box Counting)
    \end{tabular}\\

    & SP & $1.476$ & & $0.74$ & & $1.82$ & $3.75$ &  \\
 
    & & $\pm0.005$ & & $\pm 0.02$ & & $\pm 0.01$ & $\pm 0.02$ &\\
  
   Triangular &  DSP & $1.463$ & $1.47$ & $0.820$ & $1.46$ & $1.71$ & $3.69$ &
    $5.73$ \\
 
    & & $\pm 0.004$ & $\pm 0.02$ & $\pm0.006$ & $\pm0.01$ & $\pm 0.02$
    & $\pm 0.03$ & $\pm 0.05$ \\

    \begin{tabular}{p{3.0cm}p{1.5cm}p{8.0cm}}
      && $d_H=1.45\pm0.01$ (Box Counting) 
    \end{tabular}\\

    & SP & $1.466$ & & $0.76$ & & $1.91$ & $3.87$ & \\
    & & $\pm0.016$ & & $\pm 0.02$ & & $\pm 0.01$ & $\pm 0.03$
    & \\
 
    \hline

\end{tabular}
\bigskip
\caption{\label{table2} Comparison of the hull exponents and hull
fractal dimension measured for the DSP model on the square and
triangular lattices as well as with that of the SP model. $d_H^c$ is
the hull fractal dimension obtained through the proposed conjecture:
$d_H = 1 + \nu_\perp/(\nu_\parallel+\nu_\perp)$. $d_H$ is also
measured from the relation $d_H=xd_f$ considering the values of $d_f$
as $1.733$ and $1.775$ (table I) for the square and triangular
lattices respectively. Measurements are found consistent in different
methods adopted. The exponents are almost same within the error bars
on the two lattices. The hull fractal dimension in the DSP model is
found within error bar of that of the SP model.}
\end{table}

\newpage

\vfill

\vfill

\vfill

\vfill

\vfill

\vfill

\vfill

\end{document}